\documentclass[aps,prb,twocolumn,superscriptaddress,amsmath]{revtex4}

\usepackage{graphicx}
\usepackage{dcolumn}
\usepackage{bm}

\newcommand{\ra}{\rangle}

\begin{document}
\title{Work-sharing of qubits in topological error corrections} 

\author{Tetsufumi Tanamoto}
 \email{tetsufumi.tanamoto@toshiba.co.jp}
\author{Hayato Goto}%
\affiliation{%
Corporate R \& D center, Toshiba Corporation,
Saiwai-ku, Kawasaki 212-8582, Japan
}%

\date{\today}
\begin{abstract}
Topological error-correcting codes, such as surface codes and color codes, are promising because quantum operations are realized by two-dimensionally (2D) arrayed quantum bits (qubits).
However, physical wiring of electrodes to qubits is complicated, and 3D integration for the wiring requires further development of fabrication technologies.
Here, we propose a method to reduce the congestion of wiring to qubits by just adding a SWAP gate after each controlled-NOT (CNOT) gate.
SWAP gates exchange roles of qubits. Then, the roles of qubits are shared 
between different qubits. 
We found that our method transforms the qubit layout and reduces the number of qubits that cannot be accessed two-dimensionally.
We show that fully 2D layouts including both qubits and control electrodes can be achieved for surface and color codes of minimum sizes. 
This method will be beneficial to simplifications of fabrication process of quantum circuits
in addition to improvements of reliability of qubit system.
\end{abstract}
\maketitle

\section{Introduction}
Solid-state quantum computers~\cite{Yamamoto,Niskanen0,Mooij,Wallraff,Xmon1,Xmon2,Ladd,Schoelkopf} 
have made significant progress recently 
in experiments of quantum error corrections~\cite{stabilizer,Kelly,IBM}. 
Topological  error-correcting codes,  such as surface codes~\cite{Kitaev1,Kitaev2,Fowler1,Fowler2,Hill,Devitt} and color codes~\cite{Bombin,Landahl,Jones,Nigg}, 
have been intensively investigated because quantum operations are realized
through physical interactions between nearest neighboring qubits.
However, physical wiring of electrodes to qubits is complicated and 3D integration for the wiring
requires further development of fabrication technologies~\cite{Brecht}.
The complexity of wiring is mainly attributable to connections 
to syndrome-qubits from data-qubits.
Although quantum annealing machines based on superconducting qubits have already been developed~\cite{Dwave},
in order to realize quantum computation, accurate control of quantum error correction is
necessary.

Here, we propose a method to reduce the congestion of wiring to qubits.
We just add a SWAP gate~\cite{Nielsen} after each controlled-NOT (CNOT)  gate.
Because SWAP gates exchange roles of qubits, the roles of syndrome measurements are shared 
between different qubits. 
This transforms the qubit layout and reduces the number of qubits that cannot be accessed two-dimensionally.
In particular, we show that fully 2D layouts including both qubits and control electrodes 
can be achieved for surface and color codes of minimum sizes.
Because a CNOT gate is indispensable to every quantum computer,
the work-sharing of qubits by SWAP gates will be applicable to 
general quantum circuits to improve uneven workloads of qubits.

\begin{figure*}
\begin{center}
\includegraphics[width=12cm]{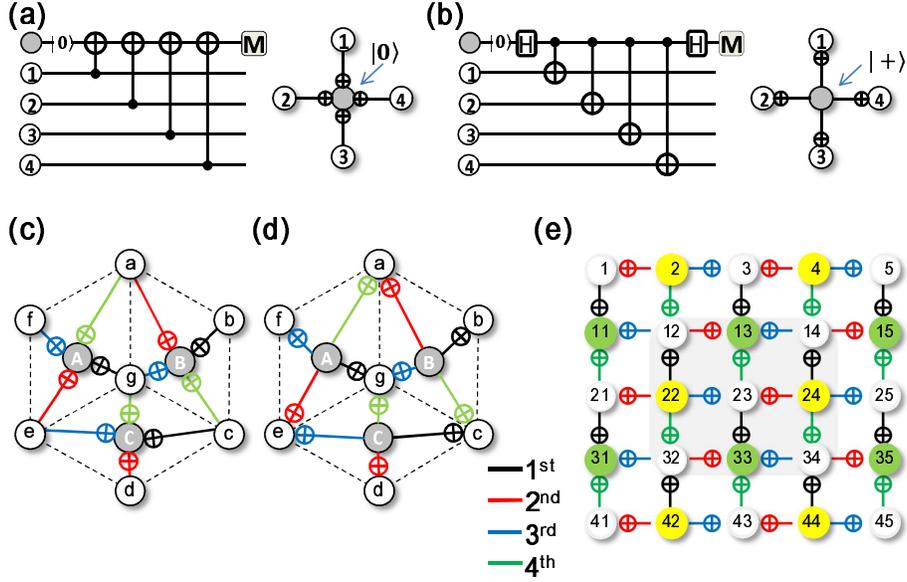}
\end{center}
\caption{
Conventional layouts of quantum error-correcting codes.
(a) A general circuit of $Z$-check for four qubit states (left) and its graphical expression (right)~\cite{Nielsen}.
Data-qubits are illustrated by open circles, 
and syndrome-qubits by solid circles. 
The qubit-qubit operations in the conventional stabilizer measurement circuits consist of CNOT gates. 
`H' indicates a Hadamard gate and `M' indicates a syndrome measurement in the computational basis. 
(b) A general circuit of $X$-check (left) and its graphical expression (right). 
(c) $Z$-checks of the 7-qubit color code~\cite{Landahl}.  `A', `B' and `C' indicate syndrome-qubits.
(d) $X$-checks of the 7-qubit color code.
(e) 
Standard stabilizer measurement procedure of a distance-3 surface code based on CNOT gates~\cite{Fowler1,Fowler2}. 
Open circles indicate data-qubits. Green and yellow circles indicates 
$Z$-check syndrome-qubit and $X$-check syndrome-qubit, respectively.
In (c)-(e), the order of CNOT operations is black, red, blue and green. 
}
\end{figure*}

Topological codes are based on the stabilizer formalism~\cite{Nielsen,Gottesman},
where a parity check process is carried out by summing  `1' of 
data-qubits by using CNOT operations.
Errors can be detected by change of the parity. 
As an example, when a wave function of four data-qubits is 
given by 
$|\Psi \ra =\sum_{i_1,i_2,i_3,i_4=0,1} a_{i_1i_2i_3i_4}|i_4i_3i_2i_1\ra$ and 
a syndrome-qubit is initialized as $|0\ra$, 
the measurement process of a $Z$-type check operator ($Z$-check) is given by $|\Psi \ra|0\ra 
\rightarrow \sum a_{i_1i_2i_3i_4}|i_4i_3i_2i_1\ra | i_4 \oplus i_3 \oplus i_2 \oplus i_1 \ra$
(Fig.1(a)), where $\oplus$ denotes summation modulo 2.
Similarly, the measurement process of a $X$-type check operator ($X$-check) is illustrated in Fig.1(b) ($|+\ra\equiv [|0\ra+|1\ra]/2$).
When the number of data-qubits increases, 
the corresponding number of wirings to a single syndrome-qubit increases.
Physical wires have a finite width and when many wires are arranged closely, 
the crosstalk problem also appears~\cite{Xmon2}. Thus, it is desirable to avoid the concentration 
of wires in a small space.

Using local interactions between nearest-neighbor qubits on a two-dimensional (2D) physical plane,
surface codes~\cite{Kitaev1,Kitaev2,Fowler1,Fowler2,Hill,Devitt} and color codes~\cite{Bombin,Landahl,Jones,Nigg} 
have a high tolerance against errors.
However, control gate electrodes 
are not generally placed in the same physical plane as qubits, 
and vertical access to qubits is unavoidable.
Figures 1(c) and 1(d) show $Z$-checks and $X$ checks of the 7-qubit color code, respectively.
The three syndrome-qubits `A-C' are connected to seven data-qubits,
where the data-qubit `g' should be accessed from the vertical direction 
by stacking an additional wiring layer.
Figure~1(e) shows a distance-3 surface code, 
where $Z$-checks and $X$-checks are performed simultaneously~\cite{Kitaev1,Kitaev2,Fowler1,Fowler2}.
The code-distance $d$ is the measure of a code in which 
$\lfloor (d-1)/2 \rfloor$ physical errors can be corrected by repeated measurements, 
and corresponds to the array size 
in the surface code.
In this code, the inner nine qubits should be  accessed vertically~\cite{Fowler1}.
Thus, many stabilizer codes lead to a concentration of 
wires to syndrome-qubits and the complexity of wiring between qubits is unavoidable~\cite{Xmon2,IBM}.
Because qubits are sensitive to decoherence, the fabrication process 
of wiring is particularly difficult in view of the state-of-the-art techniques.

Our method for reducing wiring congestion is to just add a SWAP gate after each CNOT gate in stabilizer measurements.
SWAP gates are widely used in quantum operations and exchange 
qubit states without changing system entanglement~\cite{DiV,DiV2}.
The combination of a SWAP gate and a CNOT gate (hereafter a `CNOT+SWAP gate') 
shifts the role of syndrome-qubits,
and reduces the congestion of wiring. 
The replacement of CNOT gates by CNOT+SWAP gates is effective in 
stabilizer measurements because syndrome measurements are repeated many times.
We show that this method changes qubit layouts, and importantly, for codes of small sizes, 
wiring can be set on the same physical plane as qubit array.

The insertion of a SWAP gate in each CNOT gate imposes no overhead
but rather it is advantageous
when physical interactions between qubits are $XY$ interactions~\cite{XY}. 
The CNOT+SWAP gate can be performed directly by the $XY$ interaction (iSWAP) with single-qubit rotations
(Fig.2(a))~\cite{XY,Schuch}.
On the other hand, a CNOT gate needs two iSWAP gates,
which makes the CNOT operation complicated and more fragile.
Thus, in the case of the $XY$ interaction,
the replacement of CNOT gates by CNOT+SWAP gates makes quantum operations more reliable
and saves operation time~\cite{iSWAP,iSWAP2,Wei}.
Moreover, high-precision iSWAP gates are experimentally feasible~\cite{Bialczak,Dewes,McKay}.


\begin{figure*}
\begin{center}
\includegraphics[width=12.5cm]{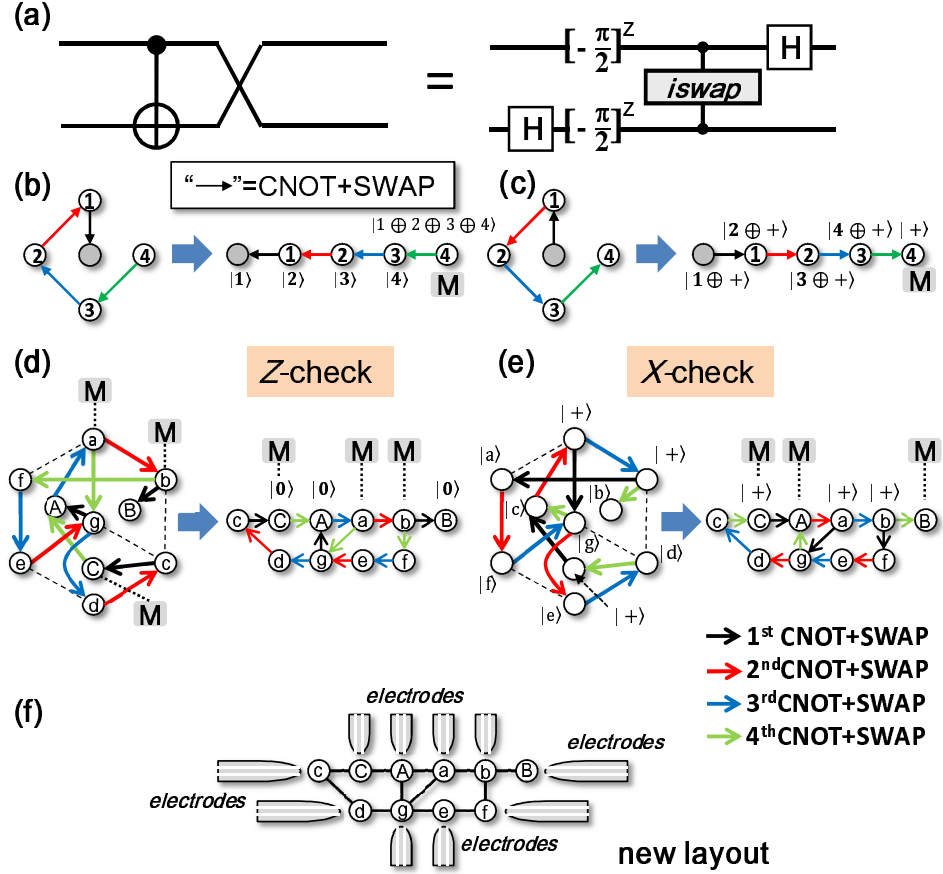}
\end{center}
\caption{
Qubit-layout transformations by CNOT+SWAP gates instead of CNOT gates. 
(a) A CNOT+SWAP gate is equivalent to an iSWAP gate~\cite{XY} with single-qubit rotations~\cite{Schuch}.
$[\pi/2]^Z$ indicates a $\pi/2$ rotations around $z$-axis.
(b),(c) Geometrical description of $Z$-check (b) and $X$-check (c) 
after the replacement of CNOT gates by CNOT+SWAP gates in Figs.1(a) and 1(b), respectively.
The order of operations is black, red, blue and green.
See equations (\ref{eqZcheck}) and (\ref{eqXcheck}).
(d) Application of CNOT + SWAP gates instead of CNOT gates 
to the $Z$-check of the 7-qubit color code (left) and its rearrangement of qubits such that connected qubits become close (right). Qubits `A', `B' and `C' are initialized to $|0\ra$ states. New syndrome-qubits are qubits `a', `b' and `C'.
(e) The $X$-check process after the $Z$-check of Fig.~(d). The syndrome-qubits  `a', `b' and `C' of the previous $Z$-checks are initialized to $|+\rangle$ states. 
After the four steps of CNOT+SWAP gates from black line to green line, qubits `A', `B' and `C' become the syndrome-qubits.
The quantum state shows that after the $Z$-check with the initialization for the $X$-check. 
(f) New 7-qubit color code layout with electrodes in the same physical plane as the qubits. 
Each electrode represents a set of wiring lines to a qubit such as 
$XY$, $Z$ and readout lines in Ref.~\cite{Xmon1,Xmon2}. 
5 qubits share the work of syndrome measurements.
} 
\end{figure*}
This paper is organized as follows:
In Sec.~II we show an example of 
the application of replacement of CNOT gate by CNOT+SWAP gate 
to fundamental stabilizer measurement.
In Sec.~III, we show layout changes of a minimum 7-qubit color 
code and the second smallest color code by our method.
In Sec.~IV, we first show layout changes 
of standard surface codes. Next we consider the application of our 
method to a rotated surface code, and we also show results of numerical simulations 
of logical error probabilities. 
Sec.~V is devoted to a conclusion. 
In Appendix, we show  detailed explanations of the numerical simulation of Sec.~IV  and 
a process of fault tolerant 7-qubit color code.

\section{Stabilizer code}
The effect of the insertion of a SWAP gate after each CNOT gate is 
easily understood when we apply this to 
the single stabilizer measurement shown in Figs.~1(a) and 1(b),
the results of which are shown in Figs.~2(b) and 2(c), respectively.
For the $Z$-check, we have
\begin{eqnarray}
\lefteqn{
|\Psi \ra|0\ra 
\rightarrow 
\sum_{i_1,i_2,i_3,i_4=0,1} a_{i_1i_2i_3i_4}|i_4i_3i_2i_1\ra |i_1 \ra } \nonumber\\
&\rightarrow& 
\sum a_{i_1i_2i_3i_4}|i_4,i_3, i_2 \oplus i_1, i_2\ra |i_1 \ra\nonumber\\
&\rightarrow& 
\sum a_{i_1i_2i_3i_4}|i_4,i_3 \oplus i_2 \oplus i_1,i_3,i_2\ra |  i_1 \ra\nonumber\\
&\rightarrow& 
\sum a_{i_1i_2i_3i_4}|i_4 \oplus i_3 \oplus i_2 \oplus i_1,i_4,i_3,i_2\ra | i_1 \ra
\label{eqZcheck}
\end{eqnarray}
We can see that the SWAP gates shift the roles of qubits one by one, and finally if we regard qubit `4' 
as a new syndrome-qubit, we obtain the same wave function $|\Psi\ra$ as that of Fig.~1(a).
As shown in Fig.~2(b), the qubit layout is now transformed to a one-dimensional qubit array.
The same thing holds for the $X$-check (Fig.~2(c)):
\begin{eqnarray}
|\Psi \ra|+\ra
&\rightarrow& 
\sum a_{i_1i_2i_3i_4} |+, i_4 \oplus +, i_3 \oplus+, i_2 \oplus + \ra| i_1\oplus +\ra
\nonumber \\
\label{eqXcheck}
\end{eqnarray}
If we measure qubit `4', we obtain the same wave function $|\Psi\ra$ as that in Fig.~1(b). 
Thus, we can avoid the congestion of wiring to one syndrome-qubit by just inserting SWAP gates. 
After the $Z$-checks in Fig.~2(b), when 
qubit `4'  is initialized to a $|+\ra$ state, and the process of 
Fig.~2(b) is reversed from the green operation to the black operation, we can restore 
the qubits to their original roles (closed circle is back to the syndrome-qubit).
Thus, serial operations of $X$-checks following $Z$-checks can be repeated,
and the role of syndrome qubit is shared between the two end qubits.

\section{Color code}
\begin{figure}
\begin{center}
\includegraphics[width=12.5cm]{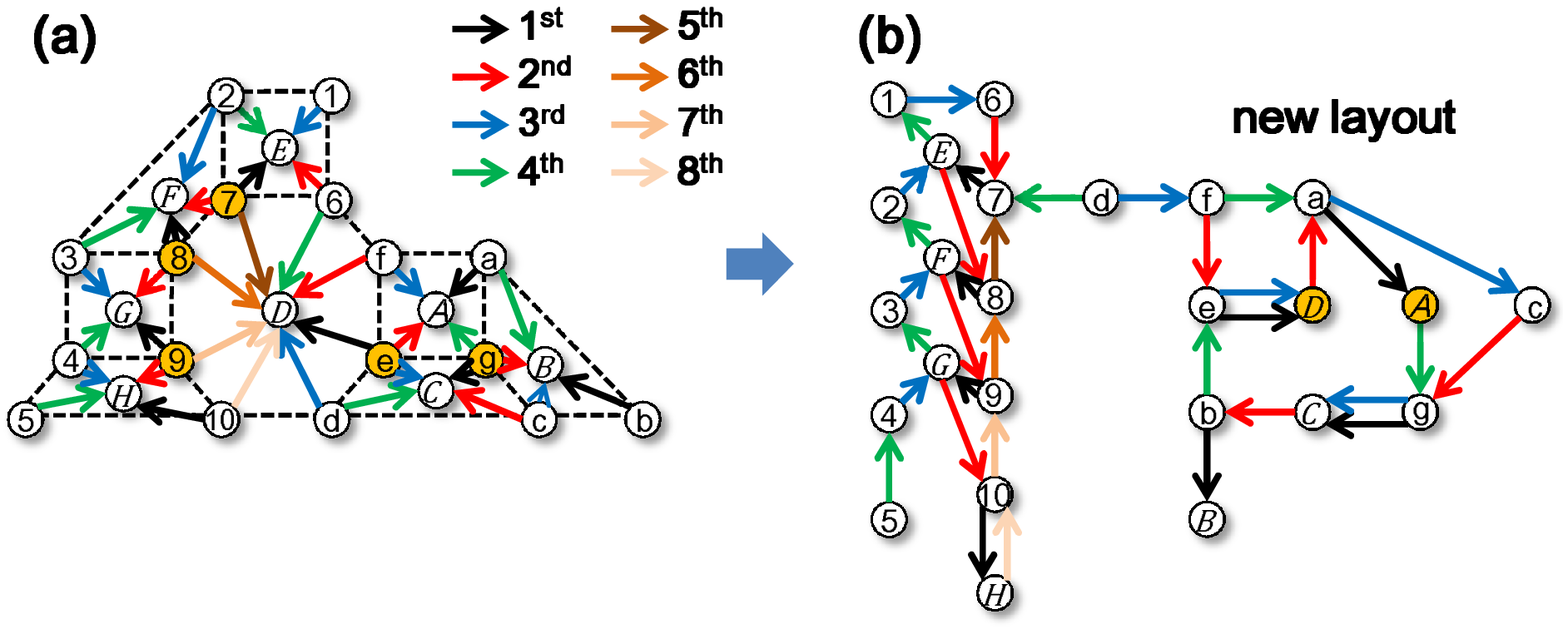}
\end{center}
\caption{
The transformation of the next smallest 4.8.8 triangular color code.
(a)  Conventional 4.8.8 triangular color code of code distance 5~\cite{Landahl}.
(b)  Replacement of CNOT gate by CNOT+SWAP gate to the 4.8.8 
triangular distance-5 color code. The transformation changes depending on 
the order of syndrome measurements. The orange circle indicates 
qubits that cannot be accessed two-dimensionally. 
By this transformation, the number of orange circles can be reduced.
}
\label{Fig3}
\end{figure}

Let us show the effect of the replacement of CNOT gates by CNOT+SWAP gates
in the 7-qubit color code,
which is the minimum color code (code distance 3)
~\cite{Bombin,Landahl}.
New connections between qubits are determined such that each step 
reproduces the same qubit states 
as those of Figs.~1(c) and 1(d).
Figures~2(d) and 2(e) show the results of applying CNOT+SWAP gates, 
instead of CNOT gates, to the 7-qubit color code.
This replacement transforms the qubit layout to a quasi-one-dimensional layout,
where 2D access to all qubits is possible (Fig.~2(f)), and
consequently no additional 3D layer for wiring is needed.
Figure~2(e) is carried out after Fig.~2(d),
where the order of CNOT+SWAP gates 
is reversed between Figs.~2(d) and 2(e). 
That is, we can use the same connections between the $Z$-checks and $X$-checks,  
and syndrome measurements, 
in which the role of syndrome-qubits is shared in five qubits (qubits `A', `B', `C', `a' and `b'), can be repeated.
The application of our method to the next larger code (code distance 5) also 
reduces the number of qubits that cannot be accessed two-dimensionally 
from 5 to 2 qubits as shown in Fig.~\ref{Fig3}.

So far, the stabilizer measurements have been
assumed to be error-free. 
Otherwise, unexpected errors propagate through 
the stabilizer measurements for the color codes.
The fault-tolerant version~\cite{Gottesman2} of the 7-qubit code is described in Appendix.

\begin{figure}
\begin{center}
\includegraphics[width=8.5cm]{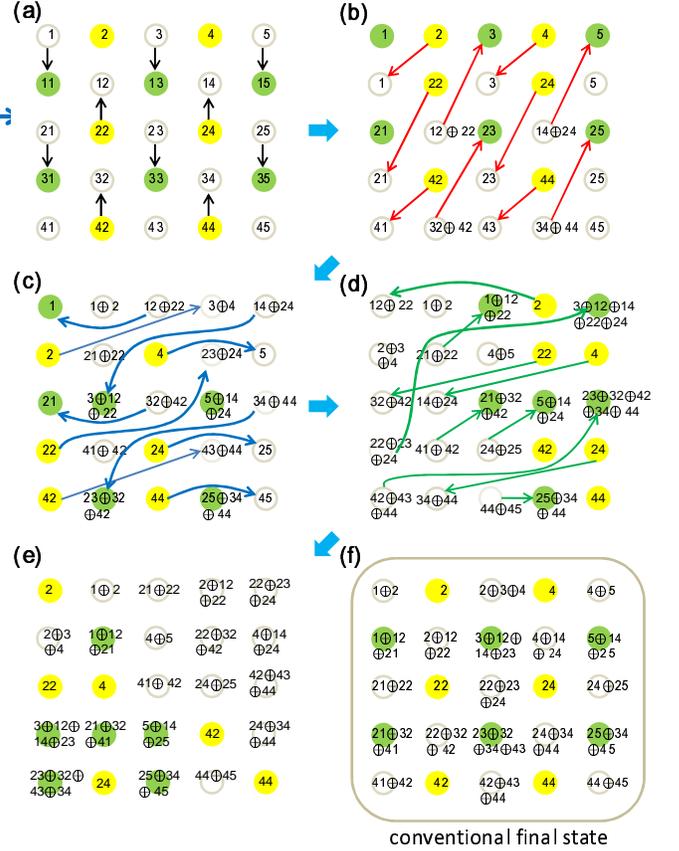}
\end{center}
\caption{
Detailed replacement process by CNOT+SWAP gate in distance-3 standard surface code.
Four directions of CNOT gates are replaced by CNOT+SWAP gates.
In each step, the CNOT+SWAP gates are applied to two qubits such that after the replacement 
quantum state is the same as that of original CNOT gate-type.
(a) 1st step. 
(b) 2nd step. 
(c) 3rd step. 
(d) 4th step. 
(e) A final quantum state of the replacement.
(f) A conventional final quantum state of Fig.~1(e) by using CNOT gates. 
The difference between (e) and (f) is the arrangement of qubits.
}
\label{surface_process}
\end{figure}

\begin{figure*}
\begin{center}
\includegraphics[width=13.5cm]{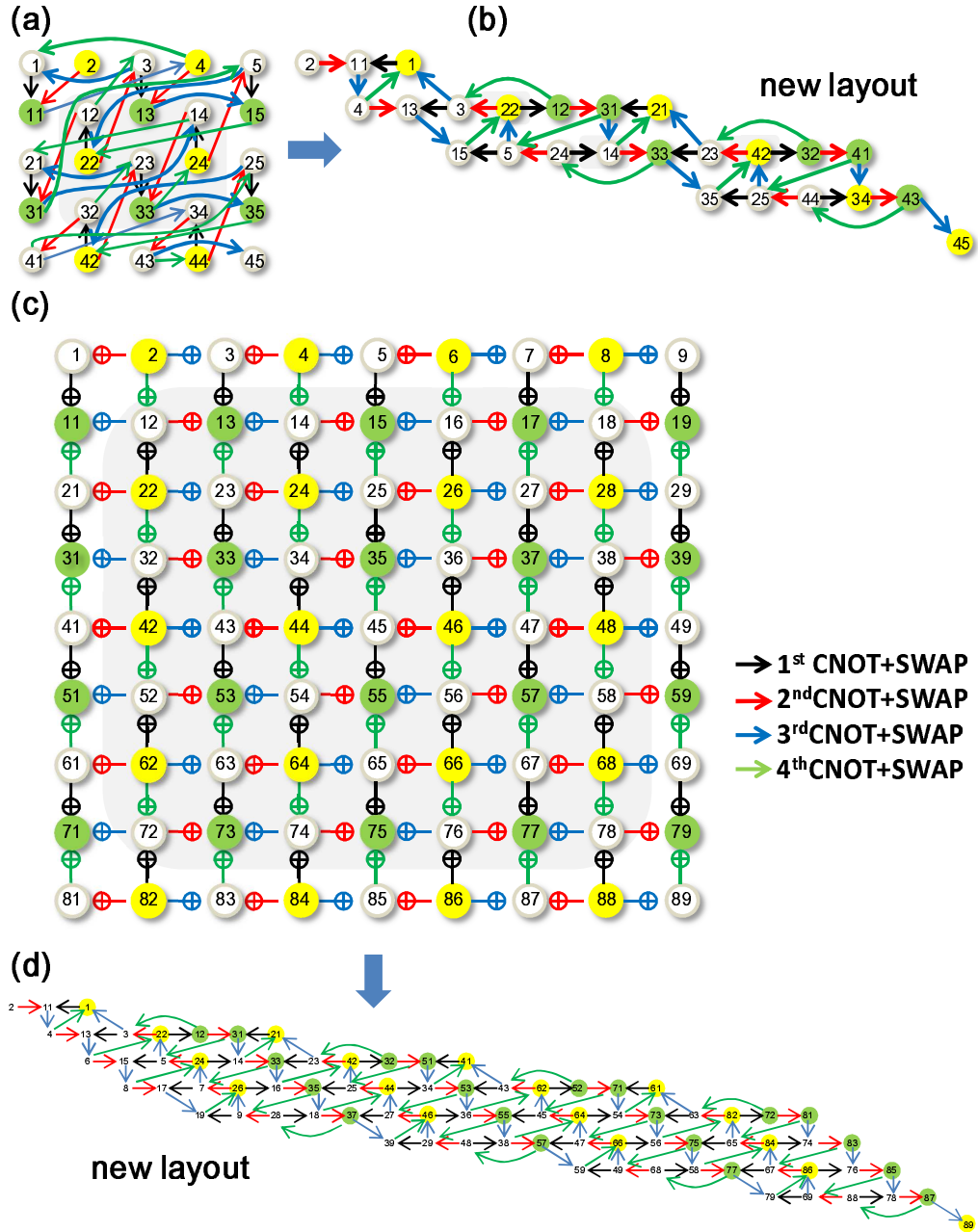}
\end{center}
\caption{
Layout transformations of surface codes. 
(a) Replacement of CNOT gates by CNOT+SWAP gates in a distance-3 surface code.
The replacement is carried out such that quantum states
after the replacement are the same as those for CNOT gates.
(b) Rearrangement of Fig.~(a) such that connected qubits become close. 
20 qubits share the work of syndrome-qubits.
(c) Distance-5 standard surface code.
(d) Distance-5 surface code
transformed by the replacement of CNOT gates 
by CNOT+SWAP gates. 53 qubits share the work of syndrome-qubits.
The order of operations is black, red, blue and green. 
} 
\label{surface}
\end{figure*}

\section{Surface code}
Next, we apply our idea to the standard surface code as shown in Fig.~\ref{surface_process}.
In each step, the quantum state of Fig.~\ref{surface_process} coincides with that of the standard state 
of Fig.~1(e). 
Figure~\ref{surface}(a) shows the new connections between qubits of a distance-3 surface code
after the replacement of CNOT gates by CNOT+SWAP gates.
When the connected qubits are rearranged closely, 
a new layout of qubits arises, as illustrated in Fig.~\ref{surface}(b),
where 20 qubits share the role of the syndrome measurement.
The number of the qubits
to which two-dimensional access is impossible 
is reduced to three (qubits `14', `22' and `42') 
from nine (=3$\times$3 in Fig.~1(e)).
Similarly, the number of the qubits that 
cannot be accessed two-dimensionally for a distance-5 surface code is reduced from 49 (=7$\times$7) to 35 (=7$\times$5),  
as shown in Fig~\ref{surface}(c) and Fig.~\ref{surface}(d), respectively. 
In general, the number of two-dimensionally inaccessible qubits for a distance-$d$ surface code 
is reduced from $(2d-3)^2$ to $(2d-3)\times(2d-5)$ by our method.

For distance-$d$ codes, a set of syndrome measurements 
is repeated more than $d$ times. 
The next processes of Figs.~\ref{surface}(b),(d) are
carried out by reversing both the order of the connections and the directions of arrows. 
That is, the green connections are carried out first and the black one last, 
reversing the directions of the interactions.
After the reversing process, the next connections are the same as Figs.~\ref{surface}(a),(c).

\begin{figure*}
\begin{center}
\includegraphics[width=12.0cm]{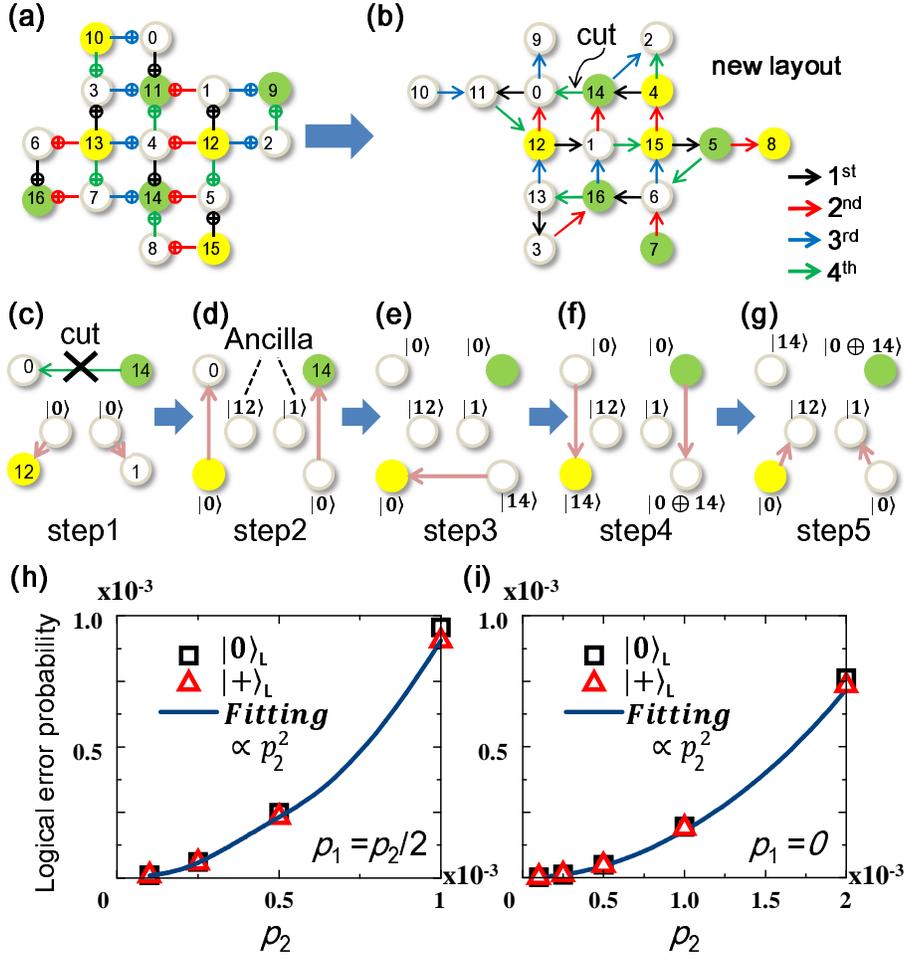}
\end{center}
\caption{
Layout transformation of a rotated distance-3 surface code.
(a) Conventional layout of the rotated distance-3 surface code.
(b) Replacement of CNOT gates by CNOT+SWAP gate 
in the rotated distance-3 surface code.
There is one qubit (qubit `1') that cannot be accessed two-dimensionally.
In order to access all qubits, we need to cut one of the connections to around qubit `1'.
Here, we show a case of cutting the connection between qubit `0' and `14'.
(c)-(g) The cut technique to connect qubit `0' with `14' using CNOT+SWAP gates.
The initialization of two ancilla qubits can be carried out during the 4-th process of Fig.~(b).
(h) 
Estimated logical error probabilities of the distance-3 rotated surface code of Fig.~(b)
as a function of physical qubit error.
The symbols indicate the logical error probabilities estimated by numerical simulation, 
where the single-qubit error probability ($p_1$) is comparable to the two-qubit physical error probability ($p_2$)
such as  $p_1=p_2/2$.
The initial states are encoded logical states of $|0\ra_L$ and $|+\ra_L$.
(i)  The estimated logical error probabilities when $p_1=0$.
} 
\label{rotated_surface}
\end{figure*}

\begin{figure}
\begin{center}
\includegraphics[width=8.5cm]{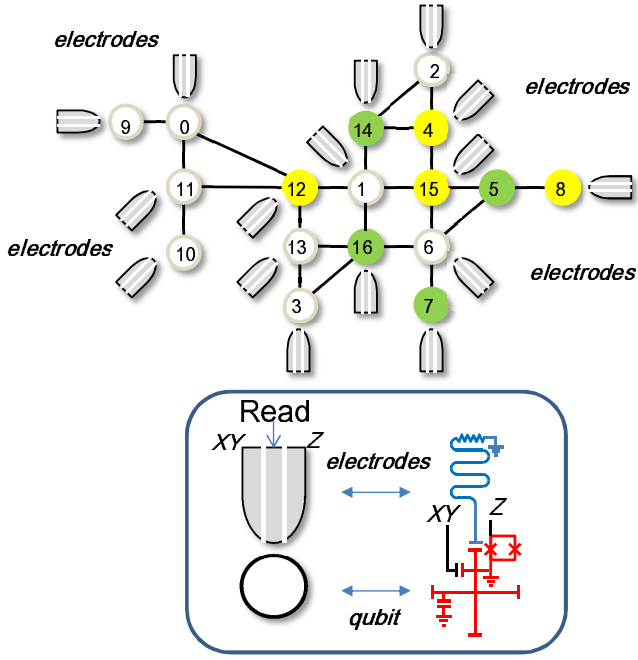}
\end{center}
\caption{
Fault-tolerant rotated surface code with electrodes in the same physical plane as qubits.
By cutting the connection between qubits `0' and `14' in Fig.~\ref{rotated_surface}(b), we can 
attach electrodes to all qubits in the same physical plane as the qubits.
The distances between qubits will be optimized to avoid cross-talk between electrodes.
Inset: Superconducting implementations 
of qubits and electrodes:
Each electrode represents a set of wiring lines to a qubit such as 
$XY$, $Z$ and readout lines in Ref.~\cite{Xmon1,Xmon2}. 
}
\label{SCelectrodes}
\end{figure}

Rotated surface codes are known to be efficient surface codes~\cite{Horsman}.
In the case of distance-3 surface codes, 
the number of qubits is reduced from 25 (Fig.~1(e)) to 17 (Fig.~\ref{rotated_surface}(a)).
The order of operations is crucial for the fault-tolerance~\cite{Tomita}.
Figure~\ref{rotated_surface}(b) shows the result of the replacement of CNOT gates by CNOT+SWAP gates. 
12 qubits share the role of the syndrome measurements.
Since the central qubit cannot be accessed two-dimensionally as before, 
we introduce a technique for cutting the connections,
as a result of which a fully 2D layout is achieved. 
The cut technique is carried out by the additional five processes shown 
in Figs.~\ref{rotated_surface}(c)-(g) with ancilla qubits. 
(These operations are performed following the fourth time step shown in Fig.~\ref{rotated_surface}(b).)  
The ancilla qubits are initialized to $|0\ra$ states and used to hold the quantum states 
of the qubits `0' and `14'.
By serial application of CNOT+SWAP gates, 
the same operations as those without the cut 
can be realized.
To demonstrate the usefulness of the cut technique,
we evaluated the logical error probabilities of the distance-3 rotated surface code
(Figs.~\ref{rotated_surface}(h),(i)) by numerical simulations~\cite{Goto} 
(see Appendix for details).
The simulation results shown in Figs.~\ref{rotated_surface}(h) and (i)
not only prove the fault-tolerance but also 
show substantial reduction of error probabilities.
New rotated surface code layout with electrodes is 
shown in Fig.~\ref{SCelectrodes} where 
each electrode represents a set of wiring lines as used in Ref.~\cite{Xmon1,Xmon2}. 

\section{Conclusion}
Replacement of CNOT gates by CNOT+SWAP gates which corresponds to  iSWAP gates 
changes layouts of quantum error-correcting codes and relaxes the congestions of wiring 
between qubits. Importantly, we showed that fully 2D layouts including both qubits and control electrodes 
can be achieved for surface and color codes of minimum sizes.
By the work-sharing, the concentration of 
workloads to specific qubits can be relaxed, and the degradation of  the qubits will be 
mitigated.  This will improve the reliability of a quantum circuit.
Moreover, it is considered that simple 2D circuits are important for 
initial development phases of experiments.
In general, it is not easy to operate circuits as expected, 
because some mistakes are often found in designs after fabrications. 
Thus, repeated fabrications of 
chips are necessary.  Simple circuit layouts are beneficial for 
a reduction of fabrication period as well as fabrication process.

\begin{acknowledgments}
 We thank A. Nishiyama, M. Koyama, H. Hieda and S. Yasuda for discussions.
\end{acknowledgments}

\appendix
\section{Simulation method.}
To show the usefulness of the cut technique, 
we performed numerical simulations and evaluated the performance. 
Figure~4(h) shows the results where single-qubit error probabilities ($p_1$) 
are comparable to the two-qubit error probability ($p_2$) such as $p_1=p_2/2$. 
Figure~4(i) shows the results 
where the single-qubit errors are rare compared to the two-qubit errors such as $p_1=0$. 
Here, we assume that the two-qubit error is an error during a 
CNOT+SWAP operation.
The curves in Figs.~4(h) and 4(i) are the fits to the simulation results with a function form of $\alpha p_2^2$, 
where $\alpha$ is a single fitting parameter. 
These excellent fits mean that any single-qubit errors have been corrected 
and therefore prove the fault tolerance of the syndrome measurements with the cut technique. 
It is also notable that the logical error probabilities are comparable to those in Ref.~\cite{Tomita}, 
where the cut technique is not used. 
Thus, we conclude that the cut technique is useful 
to realize fully accessible 2D layouts without spoiling performance.

In this simulation, we initially prepare error-free logical $|0\ra_L$ or $|+\ra_L$ 
and repeat syndrome measurements. 
In Fig.~\ref{rotated_surface}, the next process is 
carried out by reversing the order of the connections 
such that the cut is carried out first and the black one last.
At the end of each round of syndrome measurements, 
we perform recovery operations according to the measurement results,
where we estimate error positions with the lookup table decoder 
designed for the present case (see Tables 1-4). 
After the recovery, we estimate the logical error probabilities 
by error-free measurements and decoding the results. 
The error model assumed here is the standard depolarizing noise model: One of the three single-qubits 
Pauli errors occurs with probability $p_1/3$ on each idle qubit, 
after each initialization to $|0\ra_L$ and each Hadamard gate, 
and before each computational-basis measurement; 
one of the fifteen two-qubit Pauli errors occurs with probability $p_2/15$ 
after each CNOT+SWAP gate. For this simulation, 
we used the same stabilizer simulator as Ref.~\cite{Goto}. 
From the simulation results, we estimated logical error probabilities per round with the results of 40 rounds.

\begin{table}
\begin{center}
\begin{tabular}{|c|c|c|}\hline
 Syndrome 1& Syndrome 2 & Errors
\\ \hline
1 & 1 & $X_3$  \\
2 & 2 & $X_1$ \\
0 & 3 & $X_2$ \\
1 & 3 & $X_2$ \\
3 & 3 & $X_2$ \\
4 & 4 & $X_6$ \\
1 & 5 & $X_3X_6$ \\
0 & 6 & $X_5$ \\
2 & 6 & $X_5$ \\
6 & 6 & $X_5$ \\
8 & 8 & $X_7$ \\
2 & 10 & $X_5X_8$ \\
1 & 12 & $X_8$ \\
4 & 12 & $X_8$ \\
12 & 12 & $X_8$ \\
0 & 13 & $X_3X_8$\\ 
\hline
\end{tabular}
\end{center}
\begin{flushleft} 
Table 1: Lookup table for the $odd$-numbered round of $Z$-check. 
“Syndrome 1” and “Syndrome 2” 
columns show the penultimate and last rounds of syndrome measurements, respectively.
The values in the columns are $m_1+2m_2+4m_3+8m_4$,
where $m_1$, $m_2$, $m_3$ and $m_4$ are the computational-basis measurement results
of the syndrome qubits initially placed at qubit 9, qubit 11, qubit 14, and qubit 16, respectively.
\end{flushleft}
\end{table}
\begin{table}
\begin{center}
\begin{tabular}{|c|c|c|}\hline
 Syndrome 1& Syndrome 2 & Errors
\\ \hline
1 & 1 & $X_3$  \\
2 & 2 & $X_1$  \\
0 & 3 & $X_2$  \\
2 & 3 & $X_2$  \\
4 & 4 & $X_6$  \\
4 & 5 & $X_3X_6$  \\
0 & 6 & $X_5$  \\
4 & 6 & $X_5$  \\
8 & 8 & $X_7$  \\
8 & 10 & $X_5X_8$  \\
0 & 12 & $X_8$  \\
8 & 12 & $X_8$  \\
12 & 12 & $X_8$  \\
12 & 13 & $X_3X_8$  \\
\hline
\end{tabular}
\end{center}
\begin{flushleft}
Table 2 :Lookup table for the $even$-numbered round of $Z$-check. 
The values in the columns are defined as in Table 1.
\end{flushleft}
\end{table}

\begin{table}
\begin{center}
\begin{tabular}{|c|c|c|}\hline
 Syndrome 1& Syndrome 2 & Errors
\\ \hline
1 & 1 & $Z_1$  \\
2 & 2 & $Z_2$  \\
1 & 3 & $Z_1Z_2$  \\
4 & 4 & $Z_7$  \\
0 & 5 & $Z_4$  \\
1 & 5 & $Z_4$  \\
5 & 5 & $Z_4$  \\
0 & 6 & $Z_5$  \\
4 & 6 & $Z_5$  \\
6 & 6 & $Z_5$  \\
8 & 8 & $Z_9$  \\
0 & 10 & $Z_6$  \\
2 & 10 & $Z_6$  \\
10 & 10 & $Z_6$  \\
4 & 12 & $Z_8Z_9$   \\
\hline
\end{tabular}
\end{center}
\begin{flushleft}
Table 3 :Lookup table for the $odd$-numbered round of $X$-check. 
The values in the columns are $m_1+2m_2+4m_3+8m_4$,
where $m_1$, $m_2$, $m_3$ and $m_4$ are the computational-basis measurement results
of the syndrome qubits initially placed at qubit 10, qubit 12, qubit 13, and qubit 15, respectively. 
\end{flushleft}
\end{table}

\begin{table}
\begin{center}
\begin{tabular}{|c|c|c|}\hline
 Syndrome 1& Syndrome 2 & Errors
\\ \hline
1 & 1 & $Z_1$  \\
2 & 2 & $Z_2$  \\
2 & 3 & $Z_1Z_2$  \\
4 & 4 & $Z_7$  \\
0 & 5 & $Z_4$  \\
4 & 5 & $Z_4$  \\
0 & 6 & $Z_5$  \\
2 & 6 & $Z_5$  \\
8 & 8 & $Z_9$  \\
0 & 10 & $Z_6$  \\
8 & 10 & $Z_6$  \\
8 & 12 & $Z_8Z_9$  \\
\hline
\end{tabular}
\end{center}
\begin{flushleft}
Table 4: Lookup table for the $even$-numbered round of $X$-check. 
The values in the columns are defined as in Table 3.
\end{flushleft}
\end{table}

\section{ Fault-tolerant 7-qubit code.}
\begin{figure}
\begin{center}
\includegraphics[width=8.5cm]{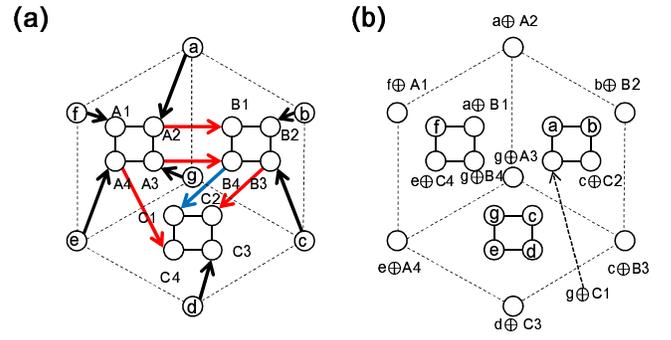}
\end{center}
\caption{
Fault-tolerant 7-qubit color code in Z-check.
Fault-tolerant 7-qubit color code is realized 
after forming three cat states as the  syndrome-qubits. 
The syndrome qubits `A', `B' and `C' in Fig.~1(c),(d) in the text are replaced 
by four qubits, `A1-A4', `B1-B4' and `C1-C4', which are assumed to be in cat states, respectively.
(a) When we can successfully obtain the cat state, additional three steps (black, red and blue arrows in the figure) of CNOT+SWAP gate 
are needed for the generation of a fault-tolerant 7-qubit color code.
(b) Final state after the operations. 
}
\label{color_FT1}
\end{figure}

\begin{figure*}
\begin{center}
\includegraphics[width=12cm]{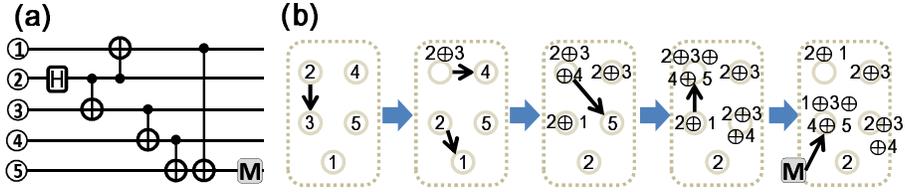}
\end{center}
\caption{ 
Generation process of a four-qubit cat state from five qubits.
A four-qubit cat state is useful for 
the syndrome-qubits of the fault-tolerant 7-qubit color code.
(a) Circuit for generating a four-qubit cat state, given by
$\frac{1}{\sqrt{2}} [|0\ra|0\ra |0\ra |0\ra +|1\ra|1\ra|1\ra|1\ra]$, 
starting from $|00000\ra$.  
(b) Generation process using CNOT+SWAP.
}
\label{cat}
\end{figure*}

One way to meet the condition of fault tolerance~\cite{Nielsen} 
is to use cat states in the stabilizer measurements~\cite{Shor}.
Figure~\ref{color_FT1}(a) 
shows a direct application of the four-qubit cat-state
$[|0\ra|0\ra |0\ra |0\ra +|1\ra|1\ra|1\ra|1\ra]/\sqrt{2}$ to the color code  in Z-check. 
(Creation of the cat state is described in Fig.~\ref{cat}.)
The connections between qubits change when it is compared with Fig.~2(d) in the text,
because at most four data-qubits can access the cat state simultaneously.
Note that the connections A4-A3-B4-C1-C4-A4 constitute a closed loop, and consequently
the qubit `g' cannot be accessed in the same physical plane.

\begin{figure*}
\begin{center}
\includegraphics[width=11.5cm]{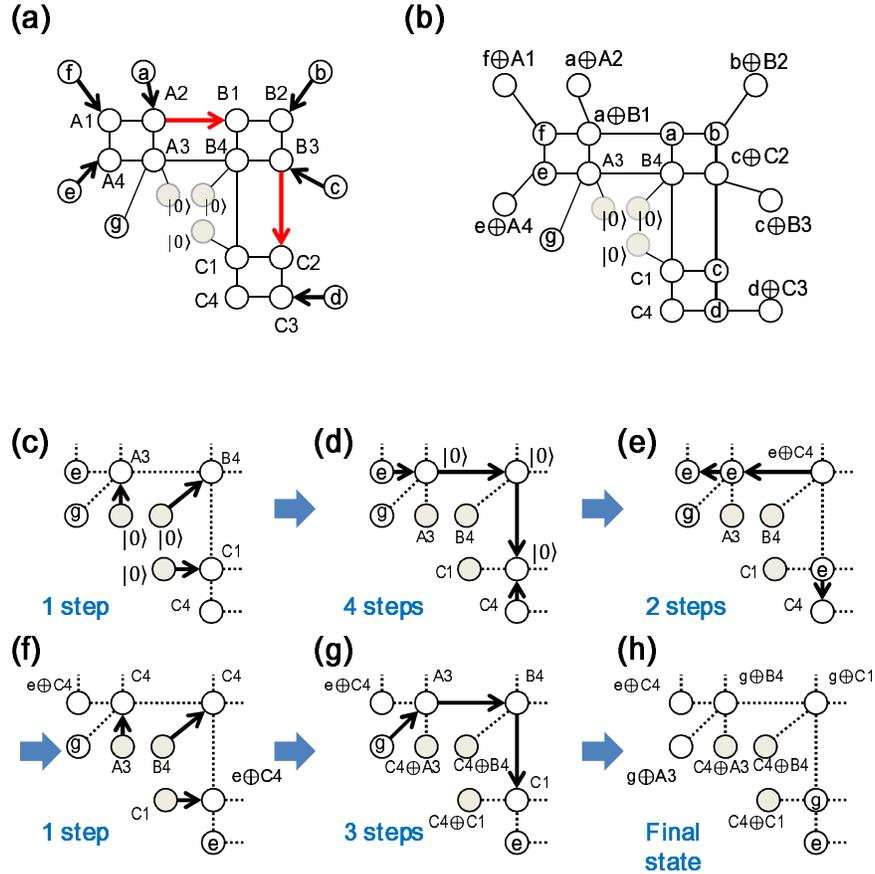}
\end{center}
\caption{
Fault tolerant 7-qubit color code, in which all qubits can be accessed \ two-dimensionally.
In order to access all qubits two-dimensionally, 
we cut the connection between  `A4' and `C4' of Fig.~\ref{color_FT1}.
The CNOT+SWAP gate between the qubit `A4' and `C4' can be established by 
a series of CNOT+SWAP gates in a A4-A3-B4-C1-C4 line, 
with additional ancilla qubits attached to the qubits `A3', `B4', `C1'
(closed dark circles). 
The three ancilla qubits are initialized to $|0\ra$ state.
(a) Operations before the connection between qubits  `A4' and `C4'.
The CNOT+SWAP gates expressed by the black arrows are carried out first, 
and those of the red arrows are carried out next.
(b) Quantum states after the processes of Fig.~(a).
(c) Apply  CNOT+SWAP gates from the three ancilla qubits to qubits `A3', `B4' and `C1'. 
Then, the three ancilla qubits store the quantum states of `A3', `B4' and `C1'. 
(d) Apply CNOT+SWAP gates from the qubit of `e' to the two nearest qubits. 
Next, apply  CNOT+SWAP gates from `C4' qubit to the above two qubits.
(e) Apply CNOT+SWAP gates to put `e$\oplus$ C4' and `e' states to their target qubits, respectively.
(f) Return the qubit states in the three ancilla to the original qubits.
(g) Apply CNOT+SWAP gates to move the qubit state `g' to the appropriate position.
(h) Final state, which has the same quantum states of  Fig.~\ref{color_FT1}(b). 
}
\label{App3}
\end{figure*}

By \textit{cutting} the connection 
between `A4' and `C4',
we can realize the fault-tolerant 7-qubit color code where 
all qubits can be accessed two-dimensionally.
Three ancilla qubits that 
interact with the qubits `A3', `B4', `C1' are introduced to store the quantum states of 
those qubits. 
The interaction between `A4' and `C4' with the cut technique is carried out 
by connecting qubit pairs among `A3', `B4', and `C1' one by one.
Additional 11 steps are required (see Fig.~\ref{App3}).

\clearpage


\begin{thebibliography}{99}
\bibitem{Yamamoto}
T. Yamamoto, Y.A. Pashkin, O. Astafiev, Y. Nakamura, and J.S. Tsai, 
Nature {\bf 425}, 941 (2003).

\bibitem{Niskanen0}
A.O. Niskanen, K. Harrabi, F. Yoshihara, Y. Nakamura, S. Lloyd, and J.S. Tsai, 
Science, {\bf 316},  723  (2007).

\bibitem{Mooij}
J.H. Plantenberg, P.C. de Groot, C.J. Harmans, and J.E. Mooij, 
Nature {\bf 447}, 836 (2007).

\bibitem{Wallraff}
A. Wallraff, D.I. Schuster, A. Blais, L. Frunzio, R.S. Huang, J. Majer, S. Kumar, S.M. Girvin, and R.J. Schoelkopf,
Nature {\bf 431}, 162 (2004). 

\bibitem{Xmon1}
R. Barends, 
J. Kelly, A. Megrant, D. Sank, E. Jeffrey, Y. Chen, Y. Yin, B. Chiaro, J. Mutus, C. Neill, 
P . \'{O}Malley, P. Roushan, J. Wenner,  T.C. White, A.N. Cleland, and J.M. Martinis, 
Phys. Rev. Lett.  {\bf 111}, 080502 (2013).

\bibitem{Xmon2}
R. Barends, 
J. Kelly, A. Megrant, A. Veitia, D. Sank, E. Jeffrey, T.C. White, J. Mutus, 
A.G. Fowler, B. Campbell, Y. Chen, Z. Chen, B. Chiaro, A. Dunsworth, C. Neill, P. \'{O}Malley,  
P. Roushan, A. Vainsencher, J. Wenner, A.N. Korotkov, A. N. Cleland, and J.M. Martinis,  
Nature {\bf 508}, 500 (2014). 

\bibitem{Ladd}
T.D Ladd, F. Jelezko, R. Laflamme, Y. Nakamura, C. Monroe, and J.L. \'{O}Brien,
Nature {\bf 464}, 45 (2010).

\bibitem{Schoelkopf}
R.J. Schoelkopf  and  S.M. Girvin,
Nature {\bf 451}, 664 (2008).

\bibitem{stabilizer}
D. Rist\`{e}, 
S. Poletto, M.Z. Huang, A. Bruno, V. Vesterinen, O.P. Saira, and  L. DiCarlo,
Nat. Commun.  {\bf 6}, 6983 (2015).


\bibitem{Kelly}
J. Kelly, R. Barends, A.G. Fowler, A. Megrant, E. Jeffrey, T.C. White, D. Sank, J.Y. Mutus, B. Campbell, 
Yu Chen, Z. Chen, B. Chiaro, A. Dunsworth, I.C. Hoi, C. Neill, 
P. J. J. \'{O}Malley, C. Quintana, P. Roushan, A. Vainsencher, J. Wenner, A.N. Cleland, and J.M. Martinis, 
Nature \textbf{519}, 66 (2015).

\bibitem{IBM}
A.D. C\'{o}rcoles, E. Magesan, S.J. Srinivasan, A.W. Cross, M. Steffen, J.M. Gambetta, and J.M. Chow,
Nat. Commun. {\bf 6}, 6979 (2015).



\bibitem{Kitaev1}
S.B. Bravyi and A.Y. Kitaev, 
arXiv:quant-ph/9811052.

\bibitem{Kitaev2}
E. Dennis, A. Kitaev, Y.A. Landahl, and J. Preskill,
J. Math.Phys.{\bf 43},  4452 (2002).

\bibitem{Fowler1}
A.G. Fowler, M. Mariantoni, J.M. Martinis, and A.N. Cleland, 
Phys. Rev. A {\bf 86}, 032324 (2012).

\bibitem{Fowler2}
A.G. Fowler, A.C. Whiteside, A.L. McInnes, and  A. Rabbani, 
Phys. Rev. X {\bf 2} 041003 (2012).

\bibitem{Hill}
C.D. Hill, 
E. Peretz, S.J. Hile, M.G. House, M. Fuechsle, S. Rogge, M.Y. Simmons, and L.C.L. Hollenberg, 
Sci. Adv.{\bf 1}, e1500707 (2015).

\bibitem{Devitt}
S.J. Devitt,
Phys. Rev. A {\bf 94}, 032329 (2016).


\bibitem{Bombin}
H. Bombin and M.A. Martin-Delgado,  
Phys. Rev. Lett. {\bf 97}, 180501 (2006).

\bibitem{Landahl}
A.J. Landahl,  J.T. Anderson, and P.R. Rice, 
arXiv:1108.5738.

\bibitem{Jones}
C. Jones, P. Brooks, and J. Harrington, 
Phys. Rev. A {\bf 93}, 052332 (2016).

\bibitem{Nigg}
D. Nigg, M. Mueller, E. A. Martinez, P. Schindler, M. Hennrich, T. Monz, M.A. Martin-Delgado, and R. Blatt, 
Science {\bf 345},  302 (2014).

\bibitem{Brecht}
T. Brecht, W. Pfaff, C. Wang, Y. Chu, L.  Frunzio, M.H. Devoret, and R.J. Schoelkopf,
npj Quantum Information {\bf 2} 16002 (2016).

\bibitem{Dwave}
T. Lanting, A.J. Przybysz, A.Y. Smirnov, F.M. Spedalieri, M.H Amin, A.J. Berkley, R. Harris,
F. Altomare, S. Boixo, P. Bunyk, N. Dickson, C. Enderud, J.P. Hilton, E. Hoskinson, 
M.W. Johnson, E. Ladizinsky, N. Ladizinsky, R. Neufeld, T. Oh, I. Perminov, C. Rich, M.C. Thom, 
E. Tolkacheva, S. Uchaikin, A.B. Wilson and G. Rose,
Phys. Rev. X {\bf 4}, 021041 (2014).


\bibitem{Nielsen}
M.A. Nielsen and I.L. Chuang,  
{\it Quantum Computation and Quantum Information}
 (Cambridge Univ. Press, 2000).

\bibitem{Gottesman}
D. Gottesman, 
quant-ph/9705052.

\bibitem{DiV}
D.P. DiVincenzo, D. Bacon, J. Kempe, G. Burkard, and K.B. Whaley, 
Nature {\bf 408}, 339 (2000).

\bibitem{DiV2}
K.M. Svore, B.M. Terhal, and D.P. DiVincenzo, 
Phys. Rev. A {\bf 72}, 022317 (2005).

\bibitem{XY}
The $XY$ model is expressed by the Hamiltonian
$
H_{xy}=\sum_{i<j} J(
\sigma_{i}^x\sigma_{j}^x+
\sigma_{i}^y\sigma_{j}^y),
$
%
where $\sigma_i^\alpha$ $(\alpha=x,y)$ are the 
Pauli matrices acting on the $i$-th qubit with 
basis $|0\rangle =|\downarrow\rangle$ and 
$|1\rangle =|\uparrow\rangle$. 
The iSWAP operation acting on 
qubits `1' and `2' is given by 
$
U_{xy}^{(12)}(t=\pi/(4J))=e^{-i(\pi/4J)H_{xy}^{(12)}},
$
such as
%
$|00\rangle \rightarrow |00\rangle$,   
$|11\rangle \rightarrow |11\rangle$,
$|01\rangle \rightarrow -i|10\rangle$, and  
$|10\rangle \rightarrow -i|01\rangle.   
$

\bibitem{Schuch}
N. Schuch and J. Siewert, 
Phys. Rev. A {\bf 67}, 032301 (2003).

\bibitem{iSWAP}
T. Tanamoto, Y.X. Liu, X. Hu, and F. Nori, 
Phys. Rev. Lett. {\bf 102}, 100501  (2009).

\bibitem{iSWAP2}
T. Tanamoto, K. Maruyama, Y.X. Liu, X. Hu, and F. Nori, 
Phys. Rev. A {\bf 78}, 062313 (2008).

\bibitem{Wei}
L.F. Wei, J.R. Johansson, L.X. Cen, S. Ashhab, and F. Nori,
Phys. Rev. Lett. {\bf 100}, 113601 (2008).

\bibitem{Bialczak}
R.C. Bialczak,	M. Ansmann,	M. Hofheinz, E. Lucero, M. Neeley, A.D. \'{O}Connell, D. Sank, 
H. Wang, J. Wenner, M. Steffen, A.N. Cleland, and J.M. Martinis,
Nat. Phys {\bf 6}, 409 (2010).

\bibitem{Dewes}
A. Dewes,  
F.R. Ong, V. Schmitt, R. Lauro, N. Boulant, P. Bertet, D. Vion, and D. Esteve, 
Phys. Rev. Lett. {\bf 108}, 057002 (2012).


\bibitem{McKay}
D.C. McKay, S. Filipp, A. Mezzacapo, E. Magesan, J.M. Chow, and J.M. Gambetta,
arXiv:1604.03076.


\bibitem{Gottesman2}
D. Gottesman, 
 Proc. Sympos. Appl. Math  {\bf 68}, 13 (2010).



\bibitem{Horsman}
C. Horsman, A.G. Fowler, S. Devitt, and R. Van Meter, 
New J. Phys.{\bf 14}, 123011 (2012).

\bibitem{Tomita}
Y. Tomita and K.M. Svore,
Phys. Rev. A {\bf 90}, 062320 (2014).

\bibitem{Goto}
H. Goto and H. Uchikawa,  
Sci. Rep. {\bf 3}, 2044 (2013).

\bibitem{Shor}
P.W. Shor,  
quant-ph/9605011. 


\end{thebibliography}
\end{document}